# Affective social anthropomorphic intelligent system

**Md. Adyelullahil Mamun[1] · Hasnat Md. Abdullah[1] · Md. Golam Rabiul Alam[1] · Muhammad Mehedi Hassan[2] · Md. Zia Uddin[3]**

## Abstract

Human conversational styles are measured by the sense of humor, personality, and tone of voice. These characteristics have become essential for conversational intelligent virtual assistants. However, most of the state-of-the-art intelligent virtual assistants (IVAs) are failed to interpret the affective semantics of human voices. This research proposes an anthropomorphic intelligent system that can hold a proper human-like conversation with emotion and personality. A voice style transfer method is also proposed to map the attributes of a specific emotion. Initially, the frequency domain data (Mel-Spectrogram) is created by converting the temporal audio wave data, which comprises discrete patterns for audio features such as notes, pitch, rhythm, and melody. A collateral CNN-Transformer-Encoder is used to predict seven different affective states from voice. The voice is also fed parallelly to the deep-speech, an RNN model that generates the text transcription from the spectrogram. Then the transcripted text is transferred to the multi-domain conversation agent using blended skill talk, transformer-based retrieve-and-generate generation strategy, and beam-search decoding, and an appropriate textual response is generated. The system learns an invertible mapping of data to a latent space that can be manipulated and generates a Mel-spectrogram frame based on previous Mel-spectrogram frames to voice synthesize and style transfer. Finally, the waveform is generated using WaveGlow from the spectrogram. The outcomes of the studies we conducted on individual models were auspicious. Furthermore, users who interacted with the system provided positive feedback, demonstrating the system's effectiveness.

**Keywords** IVA · NLP · SER · Emotion · Audio-emotion · Personal-assistant

## 1 Introduction

Intelligent Virtual Assistant (IVA) is an artificial intelligence-enabled advanced voice assistant that emulates a human and delivers personalized responses through cognitive computing and analytics. These emotion-aware conversational virtual assistants have a plethora of

✉ Md. Zia Uddin
  zia.uddin@sintef.no

Extended author information available on the last page of the article.

🖄 Springer

applications. A fine-tuned system may be employed in a variety of imaginative new ways, including voice assistants, customer service agents, robotics, therapy, animation, e-learning, or games. M. Elena et al. demonstrate, in their paper [40], the usage of intelligent virtual assistants that work as salespeople in a virtual shop, increasing customer loyalty, and satisfaction while also being accessible to the disabled. Another field that can be benefited from this technology is interactive systems for industrial assembly. The paper [6] titled "Voice-based Augmented Reality Interactive System for Car's Components Assembly" gives us a glimpse of the future voice-based AR interactive system for industrial application. This paper argues that aural language is the most intuitive tool for interaction. In their work, the authors suggest a 3D automobile assembly and disassembly system that manipulates objects using voice instructions, allowing for a simple and uncomplicated interface between the user and the system. From IBM Shoebox to Google Assistant, IVA technology has come a long way. Despite the best effort, IVA still falls short in many areas that are innate to human interaction, especially emotions and empathy. Strives are being made to clear the boundary between humans and IVA.

Most contrasting characteristics with humans are: not getting the full context of the speech, lackings in recognition of the emotions of the speaker, and no presence of personality. Current IVA can hold short conversations like asking, "what is the weather?", "do I have meetings?", "send an email", etc. Unfortunately, it is not yet possible to hold a meaningful, deep conversation. We will dissect three major limitations of current generation IVA: context, emotion, and personality.

Firstly, current IVA technology disregards context, voice tone, etc. It cannot understand if a question is serious, sarcasm, or a joke. An infamous incident took place in 2012, a man 1 killed his friend and allegedly asked "SIRI" (a commercial IVA developed by Apple) where to hide the body. SIRI answered, not getting the full context, and surely enough, the murderer followed the instruction. However, after the incident, Apple, Siri's maker, disabled this conversation, and now Siri only responds with an apology. To add to that, these sophisticated virtual agents do not know anything about languages. When we say "Alexa (a commercial IVA developed by Amazon), what is the weather today," her voice-recognition model converts speech to text based on models of digitized sounds. The reason why we get desired results is that the sentence "what is the weather today" has words like "weather" and "what," which are classified as known intent for weather report searching. If we ask, "Alexa, I do not want to know about the weather today," the response will not be an expected one.

Secondly, emotion is another region in which IVA lacks greatly. As humans are driven by emotion, there is a huge communication gap when communicating with IVA. This becomes more serious when it turns into a mental health issue. For some time now, all public-facing commercial IVA has been hard-coded to detect some trigger word. For example, if a user says, "I want to kill myself," to IVA, it will show a suicide helpline and prompt them to get help. But, if we rephrase the sentence in a different way, like "I don't want to wake up tomorrow," IVA will not show help; it will try to cancel tomorrow's schedule or do a simple search. This happens because the tone of voice is a big indicator of emotion. Without an emotional context, that sentence can mean many things, from killing oneself to sleeping on the weekend morning. An emotionally aware IVA can help a lot of people who are suffering from mental health issues.

Lastly, current generation IVA does not have a personality. It is easier for humans to relate and connect with others if they have similar personalities [39]. All of the current IVA has two significant challenges to meet: How to transmit both effective and personalized qualities in the form of a consistent and realistic speech when embedded in a computational framework [41]. The IVA needs to adapt its speech to different interaction types that users might use [33]. The benefit of having a personality in an IVA is that it can continue a proper conversation with human beings. The character of IVAs must be natural and believable and must reflect moods, personality, and expressions [48]. That is why an added personality will make a significant change in how we interact with the IVAs now, which will give us the feeling of having a companion.

The issues mentioned above are extremely complex, and there is no single simple solution to them. Researchers are attempting to devise solutions that address those deficiencies. IVAs from the current generation can be improved in various ways. Nevertheless, our research's contribution is more specific. We have developed a novel, emotion-aware IVA that can transfer any person's voice style given, we feed the person's sufficient audio voice.

The key contributions to this research include:

- A social anthropomorphic intelligent system has been proposed that can classify proper emotions With Parallel 2D CNN and Transformer-encoders.
- A context retrieval technique is proposed that transcribes the voice to text using DeepSpeech while also generating appropriate emotional conversation responses with Blender from both classified emotion and transcribed text.
- Finally a voice synthesizer model is proposed that generates proper affective audio responses with personality traits like tones/cues in the synthesized voice with transferred style, using Flowtron.

## 2 Literature review

Speech Emotion Recognition is not a new research interest. We can find a conference paper [12] dated back to 1996, which tries to incorporate emotion with speech using statistical pattern recognition. Even though research interest in software emotion recognition has not changed for decades, its methodology certainly did.

The traditional approach to recognizing emotion from speech comprises Modeling, Annotation, Audio Features, and Textual Features [5, 17, 55, 63]. According to L. Devillers et al., the most crucial thing after a model is to collect the appropriate emotional audio dataset that is well labeled and suitable for a model that focuses on emotion representation [15]. As B. W. Schuller et al. explain in their article [54], the Observer Rating may be an appropriate label in automated emotion detection since it focuses on what emotion the speaker conveyed to the dataset rather than what emotion the speaker felt. Many prior research employed acting or targeted elicitation to avoid rating annotations.

Because of the limits of the old approach, scientists and researchers are turning to more recent techniques like Deep Learning. However, some of the conventional SER's intrinsic limitations remain in the new methodology. Before we begin, we must first explore the SER challenges and how scholars attempted to address them. We will encounter three key challenges.

First and foremost, a holistic speaker model is required for accurate emotional identification, taking into consideration vocal characteristics such as tone, pitch, loudness, speech speed, and voice condition. The acoustic environment, exhaustion, drunkenness, hoarse voice due to a cold, age of the speaker, accent, and a variety of other circumstances can all have an influence on the speaker's tone. In their study, B. Aazam, A. Dariush, and H. Mahdi show that a speaker's age can be predicted with reduced mean absolute error using audio speeches [1]. This demonstrates that a speaker's age influences the type of speech spectrogram he or she creates. Deviation of certain states and traits has been proven to affect model performance [20, 44]. To overcome this problem, several DNN techniques have been proposed. Glorot et al., in their research [20], demonstrated that Stacked Denoising Auto-Encoders could extract audio features without labeled data or human supervision. With the aid of a simple rectifier unit, this is achievable. Furthermore, their method greatly enhanced generalization over the baseline. Similarly, Deng et al., in their study [13], deal with the scenario where training and test corpora come from separate datasets. To decrease the disparity and extract common features created by multiple states, such as speakers, environment, and language, they suggested a "shared hidden-layer autoencoder" approach. The results of the experiments revealed that it outperformed other domain adaptation models. S. Mohammad and B. Azam presented [38] a minimum vector of features for identifying the state of emotion in a speech in their study. The proposed vector is highly optimized for minimizing computing resources. To create fourteen feature vector components, the model integrates prosodic and frequency characteristics such as MFCC, energy, and fundamental frequency. The author's subsequent work [53] combines the previously described optimized feature vector with the Hidden Markov model, which not only streamlines the technique but also improves performance and overall accuracy. Because real-time applications necessitate speed and constraint processing, such an optimized technique might be employed in a multimodal system to recognize the speaker's mood in real-time and synthesize the answer. By minimizing the feature vector and making the resulting audio-feature spectrogram image more distinct, we may be able to improve the prediction accuracy of CNN-based model.

Secondly, an emotion recognition system's success depends on effective data collection. Since its inception, there has been a scarcity of audio speech data that has been accurately categorized with the emotional states expressed by the audio speech. In the recent past, attempts have been made to collect or create audio speech data and identify it accurately [20, 44, 63]. Weakly supervised and semi-supervised techniques have been devised to address this data scarcity. After training a model, semi-supervised learning algorithms [14, 34] could correctly label the rest of the dataset. According to S. Mohammad and B. Azam, when a computer program is attempting to identify the emotional state of speech, the meaning of the words has less of an influence on the detection of the emotional state [36]. As a result, the statements uttered in the audio voice must be neutral in order for the classifier model to be free of bias from emotive phrases. The statements included in one of our datasets (RAVDESS) were of equal syllable length, and the word frequencies and familiarities were matched using the MRC psycholinguistic database. In addition, another dataset named "TESS" employed emotionally neutral words to build its dataset. In different emotional states, 200 target words were uttered with the line "Say the word _," with the target word in the blank space. As a consequence, the audio speech remarks delivered were neutral. To keep the data quality up, we just cannot rely on machine-generated labels. It is better to keep humans somewhat involved in the labeling process. This hand-to-hand cooperation of human labeling with a machine (Semi-supervised learning) is called Active Learning. On the other hand, anyone may create linguistically and emotionally matched speech audio data and compare it to the original reference audio speech to filter out the less similar ones [54]. In their work, B. Azam, A. Dariush, and N. Sadegh proposed a unique approach [7] for separating audio data from music for the Persian language, based on pitch and characteristic-based detection. Music is extracted from the classified vocal and non-vocal components of the speech using the pitch criterion and the properties of the cepstral coefficients. The classification technique using the audio features and phase selection is a radical approach that demonstrates the total superiority of the new method for the accurate separation of speech from music. This technology might be used in the future to filter away background music and just extract the necessary voice audio in scenarios where the user might be surrounded by music.

Generative adversarial networks are also used to generate audio speech and contrast it to the original reference audio. Two neural networks are used to accomplish this. The first neural network attempts to create speech in the form of audio. Simultaneously, the second neural network evaluates the difference and determines which sample is original and which is created [9]. Furthermore, transfer learning can help to reduce the disparity between produced and original utterances. Transferring sentiment from text to image has been achieved via transfer learning [69]. In SER, we feel a similar strategy may be applied. Recently, A. H. M. Seyyed, B. Azam and R. K. Mohammad introduced TRCLA, a method to address a new challenge in Transfer Learning: negative transfer. The approach is a cellular learning automata-based (CLA) transductive learning algorithm [57]. In the case of negative transfer (NT), two additional decision criteria: merit, and attitude parameters are presented to CLA. These changes led to improved performance, outcomes, accuracy, and convergence rate for Transfer learning. TRCLA's thresholding mechanism is the NT's limiting step. Using transfer learning, this efficiency can be applied to voice emotion recognition. In their paper, Google introduced one of the most current advances in the field of transfer learning. The suggested system includes a speaker encoder network, a seq-to-seq synthesis network based on Tacotron 2, and an autoregressive Wavenet-based vocoder network. The authors removed batch normalization from their Tacotron's text encoder and chose instance normalization instead. They employed a decoder-enabled neural network to eliminate Tacotrons' key features, which comprise two layers termed Prenet and Postnet. In the Singing Voice Synthesis Paper, an attention mechanism based on "tanh" was applied [42]. Researchers employed a speech synthesis model called Flowtron to synthesize spoken audio samples. The mean opinion score shows that the produced samples were quite comparable to the reference audios. The value was similar to that of other cutting-edge voice synthesis models.

Finally, the Naturalness of Generated Speech Emotion is of concern. Emotion generation remains a challenge due to its ambiguity and inherent complexity. Although very little has been done in this area, there are a few studies that attempt to address the issue. While contrasting two voice synthesis methods: Unit selection (data-based approach) and Statistical parametric (process-based approach), S. Mohammad and B. Azam addresses the gap between the naturalness factor of synthesized speech [37]. In human-machine communication, such as speech, statistical models represent the distribution of parameter values. The statistical parameters are predicted using a robust statistical model, usually HMM (hidden Markov model). In their research [32], Lee et al. present an end-to-end mode voice synthesizer using context vector and residual connection at recurrent neural networks that can construct emotion given emotion labels. Another noteworthy paper [4] is by Akuzawa et al., who employ VoiceLoop, an autoregressive SS model, in conjunction with Variational Autoencoder (VAE) to overcome the absence of global characteristics of speech limitation. When compared to VoiceLoop without label and control speech expression, this upgraded technique produces higher quality speech. Despite all efforts, voice generation is far from being natural due to the intrinsic intricacy and non-linear structure of emotion. In terms of producing samples that are more intelligible to humans, generative adversarial networks have shown potential [11, 66].

However, we want to illustrate the importance of the Generative Adversarial Network (GAN) in the SER field. Numerous GAN versions have been developed by different researchers and have proven effective in many real-life settings since the commencement of GAN [21] in 2014 by Goodfellow et al. There have not been as many, but some attempts have been made to integrate GAN with audio synthesis and generation. In their study [47], Pascual et al. suggest a generative adversarial framework with an end-to-end speech enhancement method, which is a viable alternative to the existing methodology that operates in the spectral domain and utilizes some higher-level features. The model's encoder-decoder fully-convolutional structure allows it to work quickly on denoising waveform chunks. VoiceGAN [19] is a unique neural network model for speeches that can create human-like vocal audios. Instead of focusing on what the speaker is saying, it trains to replicate the target speaker's vocal characteristics and create mel-spectrograms. The spectrograms are then translated into the time domain for voice audios using the Griffin-Lim technique. WaveNet [43], proposed by Oord et al., turns mel-spectrograms into time-domain audio waves that humans hear as speeches. In the publication Parallel WaveGAN [68], they suggested a technique for producing audio waves using the generative adversarial network method. WaveNet is trained in this approach such that current mel-spectrograms are independent of past time slices. In addition, adversarial losses are reduced as a result of the training. One of Google's studies in the field of text to voice models presented "Tacotron" [67], which synthesizes speech directly from characters. This is an end-to-end generating text-to-speech model. Furthermore, in another publication [68], Google researchers present a system that uses a recurrent neural network to predict the next feature based on prior features and context, as well as generate Mel-spectrograms with regard to character embeddings. Using vocoder, a modified wavenet [58], these Mel-spectrograms are transformed into a time-domain waveform. The combination of these two processes produces high-quality audio speech from text transcriptions and is referred to as "Tacotron 2." Recently, MelGAN [31] overcomes earlier model limitations by making architectural improvements and simplifying training methodologies. This model does not predict Mel-spectrograms based on prior Mel-spectrograms and uses a convolutional technique, which results in fewer parameters than other competing fully connected neural networks models. One of the most significant benefits of this model over rival models is its quick training speed.

In the recent past, we have seen a new model architecture that has yielded outstanding performance in the field of speech emotion recognition by utilizing a convolutional neural network approach with a long short-term recurrent neural network added at the last layer [63]. On the other hand, researchers presented Glow [29], which employs an invertible 1 x 1 convolution to produce a rudimentary form of generative flow. Using this convolution, they have improved the "log-likelihood" of any data on established benchmarks. It will also be employed in speech-generation models such as WaveGlow [51], which will create high-quality speech from Mel-spectrograms. WaveGlow generates high-quality audio waves by combining WaveNet [43] and Glow [29] model architectures.

Conversational agent development has experienced a massive shift in paradigm latterly. Google introduced Meena [2], an open-domain chatbot powered by a massive 2.6B parameter neural network model, in early 2020. The research suggests a technique to demonstrate human-level logic by increasing the likelihood of the next token based on massive amounts of conversational data gathered from social media. In a newly released study [18] by Facebook AI, they demonstrated different versions of conversational bots with many parameters. These models include the characteristics of a real human-like dialogue, such as having a personality, providing adequate time to the writer, having knowledge as well as empathy, raising various topics to maintain a healthy conversation, and so on.

The SER system's reliability may be evaluated by reading papers in which academics discuss recent challenges in the associated domain. However, a shortage of correctly labeled data, as well as different views of the same labeled data, are posing challenges to achieving the requisite accuracy in the speech emotion detection domain.

## 3 Dataset description

We used multiple datasets to get well-rounded data for our models. Some of the data were created professionally, and some were crowdsourced. We were careful not to introduce any bias in our model, so we tried to balance every data with an equivalent counterpart. For example, SAVEE had only male actors, so to compensate, we have added a TESS dataset that contains only female actors. To ensure we get an accurate representation of the real world, we have added the CREMA-D dataset, which is very diverse and contains audio with different accents and quality. We have added a summary of the datasets we have used below.

### 3.1 RAVDESS

RAVDESS [35] is the short form of "Ryerson Audio-Visual Database of Emotional Speech and Song." Although the full dataset contains speech and song, audio, and video, we will be only using the speech audio-only files (16bit, 48kHz .wav) for our purpose. This speech audio dataset consists of 1440 files (60 Trials x 24 Actors), which were done by professional actors (12 female and 12 male) vocalizing two lexically similar statements in a North American accent. The speech dataset includes neutral, calm, happy, sad, angry, fearful, surprise, and disgust expressions with two levels (normal, strong) of emotional intensity.

### 3.2 SAVEE

The full meaning of SAVEE [61] is Surrey Audio-Visual Expressed Emotion. This dataset has high-quality audio of only male voices. There are four native English male speakers who are from the University of Surrey. The use of this male-only dataset will create biases in the models that will be trained. That is why it is advised to use this dataset with other datasets with more female (in our case, we used TESS) speakers. There are seven emotional categories of data in this dataset: anger, disgust, fear, happiness, sadness, surprise, and neutral. The age of the male voices was from 27 to 31 years. The text material consists of 15 TIMIT sentences for each emotion. For one emotion, there are three common, two emotion-specific, and ten generic sentences that were different for each emotion and phonetically balanced. The three common and 12 emotion-specific sentences were recorded as neural to give 30 neutral sentences. In Total, there are 120 utterances per speaker.

### 3.3 TESS

TESS [49] is the short form of the Toronto Emotional Speech Set. This dataset consists of the voices of two actresses aged 26 and 64. As a whole, there are a set of 200 target words that were spoken in this dataset. The audio recordings resemble each of the seven emotions: anger, disgust, fear, happiness, pleasant surprise, sadness, and neutral. There are 2800 audio files of the wav format in which the two actresses uttered 200 target words with respective emotions. It is recommended to use this dataset with male-only datasets to avoid biases in the generated model.

### 3.4 CREMA-D

CREMA-D [10] stands for "Crowd-sourced Emotional Multimodal Actors Dataset." This is a dataset of 7442 audio clips from 48 male and 43 female actors between the age of 20 and 74. The actors come from different races and ethnicities like African American, Asian, Caucasian, Hispanic, and Unspecified. This is the most diverse dataset of all the datasets we have included in this paper. The actors are assigned to speak from a selection of 12 sentences using one of six emotion categories. The emotion categories were Anger, Disgust, Fear, Happy, Neutral, and Sad. They had four levels of intensity (Low, Medium, High, and Unspecified). The participants rated the emotion and emotion levels judging from audio-only, video-only, and audiovisual presentations. The process was crowdsourced, and a total of 2443 participants each rated 90 unique clips consisting of 30 audio, 30 visual, and 30 audio-visuals. 95% of the clips have more than seven ratings.

### 3.5 Indic TTS

Indic TTS [27] is a project that uses a consortium of a high-quality corpus for building text to speech synthesis systems for 13 major Indian languages [8], which includes Bengali too. The dataset includes audio speeches along with text transcriptions. For each primary language, there is a male and a female speaker who utter the lines from various domains such as newspapers, fiction, science, etc. Moreover, audio speeches are recorded in a quiet and echo-less environment. The sampling rate of the recorded audio signals is 48KHz. The recorded audio data uttering English sentences in Bengali accent and Hindi accent of both males and females is 15.23 hours and 15.75 hours. This speech corpus [8] is intended to create various speech synthesis systems in the Indian language and English, where the systems will work better for the Indian accent.

### 3.6 LJ speech

LJ speech [26] is a dataset created by Keith Ito and Linda Johnson. This is an entirely public domain dataset. One speaker who is Linda Johnson herself uttered 13100 short audio clips from 7 non-fiction books. The total length of audio clips is almost 24 hours. English transcription is created for each audio clip. Also, the audio clips are not fixed in length, varying from 1 second to 10 seconds. The dataset authors manually matched text transcription to the audio, and a QA was passed to prove that the transcripted words correctly matched with the audio speeches.

### 3.7 Libri TTS

LibriTTS [70] is an extensive dataset totaling 585 hours of audio speeches of the English language. This multi-speaker English Corpus is created for building Text-to-Speech models and further research in this field. The audio signals sampling rate is 24kHz for this dataset. This dataset is generated from another corpus called LibriSpeech [45], changing the original dataset's different characteristics. The changes include changing the sampling rate to 24kHz, adding contextual information, excluding background noises, and including original and normalized texts in the dataset.

## 4 Features & augmentation

### 4.1 Features

Different data cleaning procedures like noise removal, making the audios of equal lengths, and equally padded with silence at the beginning and end of the audio clips have been done with our datasets. We need the correct data and a good representation of our data for classification and predictive models, which we call features. We have identified multiple features of our audio data that we used to feed different models to experiment and get better results.

#### 4.1.1 Short-Time Fourier Transform (STFT)

Short-Time Fourier Transform (STFT) is the baseline of all the features that we are going to discuss. STFT divides the audio waves into different equal segments, which are short and overlapping. After that, the Fourier transform of each segment is used to generate power spectrograms. The goal of making power spectrograms is to identify resonant frequencies in the waveforms. The advantage we get from doing an STFT is that it identifies the changes in the audio signals in time series data.

#### 4.1.2 Mel-spectrogram

The Mel-spectrogram is a Mel-scale representation of frequencies created by fast Fourier transformation. The audio wave signals are converted from the time domain to the frequency domain by a short-time Fourier transform using short and overlapping segments over the audio signals, and this is called the spectrogram. The spectrogram's frequency axis is then converted into a log scale as we humans have a minimal range of recognition of frequencies and amplitudes. Also, the color dimension is converted to decibels. Finally, the frequency axis is mapped on the non-linear Mel Scale to generate a Mel-spectrogram. Mel-spectrograms are simplified analog representations of the power spectrograms in the Mel-frequency scale. This is another feature that can be used in different classification models (Fig. 1).

#### 4.1.3 Mel-Frequency Cepstral Coefficients (MFCC)

Mel-frequency cepstral coefficients identify the changes in the pitch of audio signals. It is a mathematical function to transform power spectrograms of an audio signal generated by STFT into a small number of coefficients, representing the power of that audio signal in the frequency domain. There are some mathematical procedures that are done one after another for this transformation. First, STFT is used to generate audio power spectrums. Then, frequency bins are generated by applying triangular, overlapping window functions to the power spectrograms and taking the sum of each window's energy. After that, the frequencies of the audio signal's power spectrograms are mapped in the Mel Scale (Fig. 2).

This mapping helps to finalize the number and position of window functions and the width of the frequency bins. The reason for using Mel Scale is that humans hear the audio pitches based on frequency ratios, and it is a non-linear pitch scale that represents the audio pitches in "mels" of audio in terms of its frequency. Window functions and frequency bins altogether are called mel filterbanks. Then, the log of the sum of audio signals power spectrogram, also known as cepstrum, is taken for each filterbank. Finally, for each filterbank, discrete cosine transform (DCT) is applied to the log of the sum of the power spectrograms to decorrelate them since there are correlations between filterbank energies. The benefit of using a discrete cosine transform is that it generates coefficients so that the audio signal is fairly represented by only the top few coefficients. So, the amplitudes of the discrete cosine transform of the log of the sum of the filterbank powers with respect to time are mel-frequency cepstral coefficients. MFCC paves us the way to deconvolutionize audio signals to identify resonant frequencies.

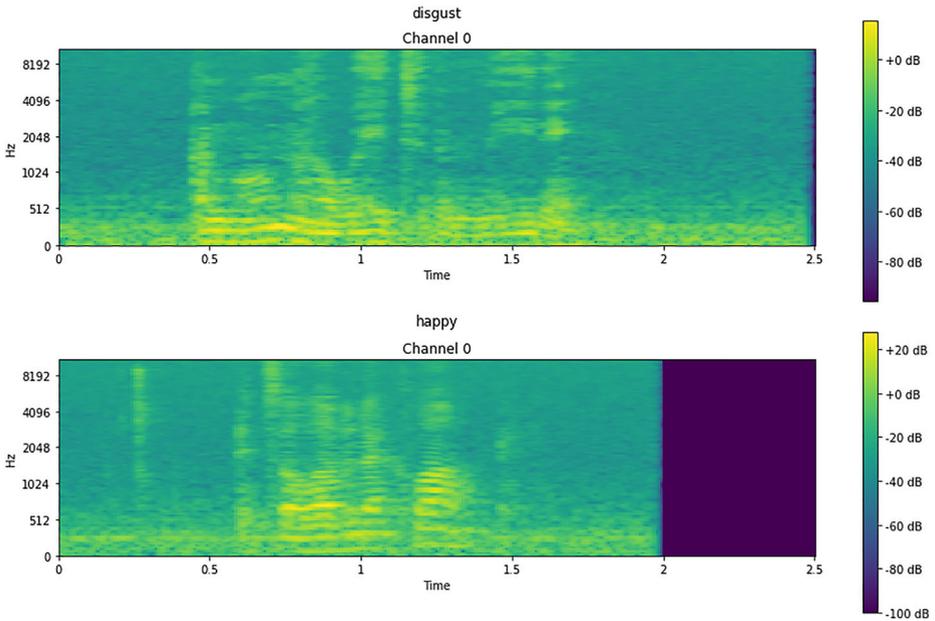

**Fig. 1** Mel-Spectrogram (Sad, Angry)

### 4.1.4 Delta

Delta is the derivative of coefficients. In other words, Delta gives us an overview of the changes in coefficients (Fig. 3). It helps us to identify the audio speeches better. With respect to time, the Delta of MFCC will represent a better understanding of the dynamics of power spectrums of audio signals. We will be using MFCC with the Delta of MFC coefficients combinedly as a feature for our models.

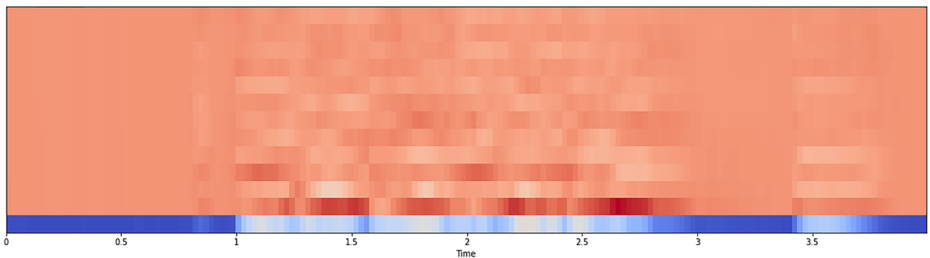

**Fig. 2** Mel-Frequency cepstral coefficients

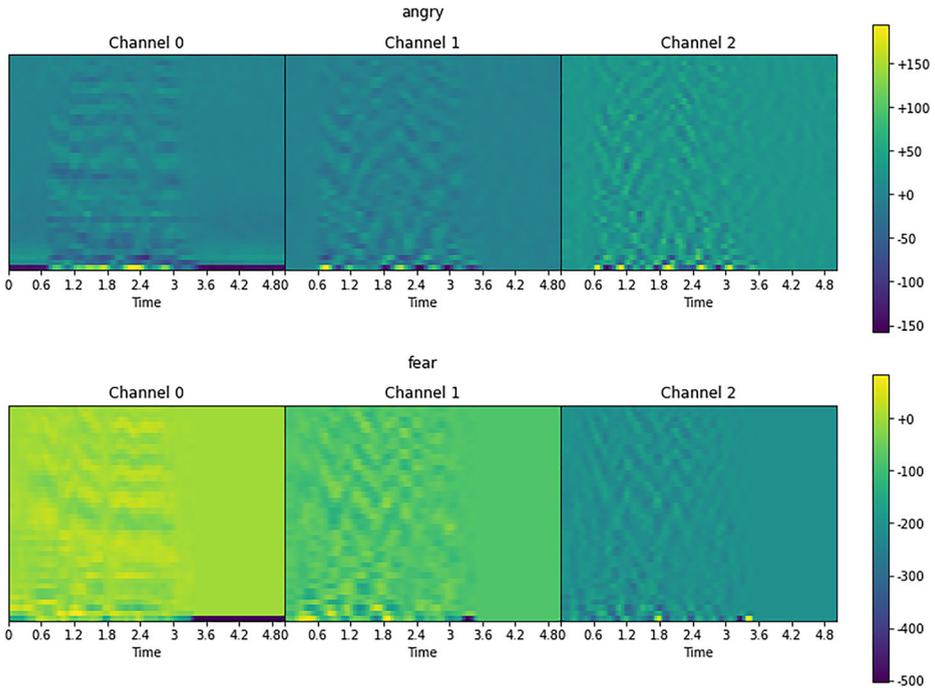

**Fig. 3** Delta

## 4.2 Augmentation

### 4.2.1 Add noise

Adding noise to the audio signal data can help the machine learning models to generalize the function better. For audio emotion recognition models, adding noise to the dataset can give the model an edge for better accuracy. This is how a typical audio speech sample looks before and after adding noise such as Additive White Gaussian Noise with a sample audio speech:

### 4.2.2 Signal Loss

Recording audio can suffer from loss of signals in the natural environment due to different hardware and latency issues. Most of the audio datasets are created in a noise-free environment for the clarity of the data. The machine learning models need to perform better in natural environments too. That is why the dataset it is training on should resemble characteristics of the natural environment. Hence, signal loss is applied to audio signals for augmentation (Fig. 4).

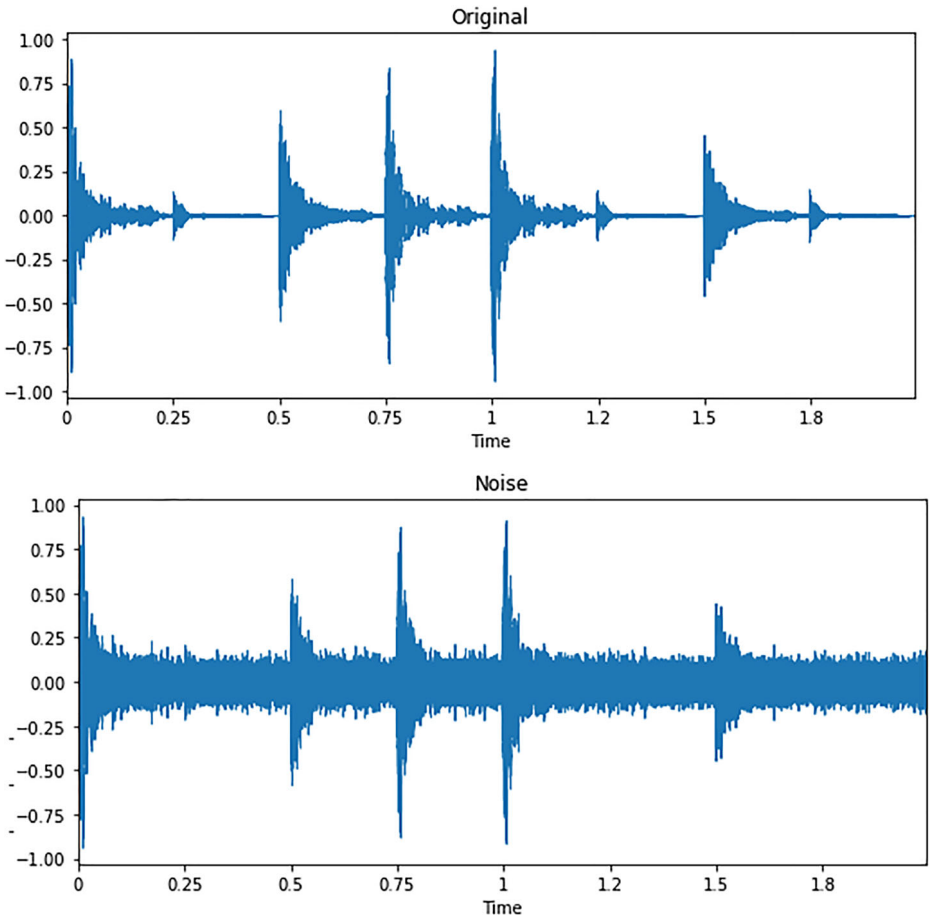

**Fig. 4** Noise Augmentation

### 4.2.3 Change volume

Generally, well-curated audio speech datasets maintain a steady level of volume. The people who create audio datasets are given proper rest to lessen their fatigue from long hours of audio sessions to maintain the same level of tone throughout the recording sessions. On a real-life environment, people do not talk like we trained actors with the systems. Sometimes they talk loudly. If the machine learning system is not robust to the loudness of the speech or environment, it can perform inaccurately. That is why for augmentation, we seldom change the volume of some data just for the machine learning model's generalization purpose.

### 4.2.4 Spectrogram augmentation

Google has introduced a new augmentation method called SpecAugment [46] for automatic speech recognition. Conventional augmentation procedures are done over the audio signals. Google's SpecAugmentation applies the augmentation process on the spectrogram of the audio which is an image representation of the signal (Fig. 5). This method does not cause any additional computational cost or data like other augmentation methods. After creating the spectrograms, SpecAugment distorts the spectrograms in the horizontal and vertical axis which are respectively Mel-frequency channels and time steps. This augmentation technique gives the neural network models generalization above the loss of information in the speech signals.

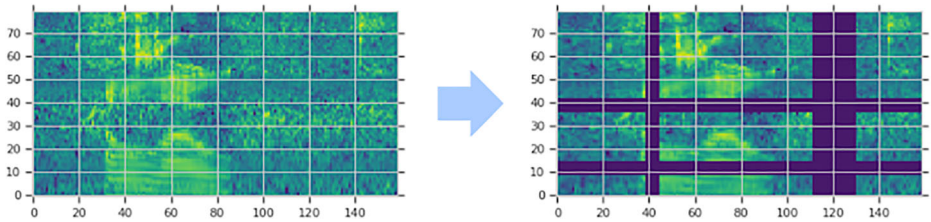

**Fig. 5** Spectrogram augmentation

## 5 Methodology

As shown in Fig. 6, speech audio signals are taken as input from the user, which can be any command or conversational speech. The system analyzes the audio signals and extracts various useful features like Mel-spectrogram, Mel-frequency cepstral coefficients (MFCC), and Delta of coefficients. These features work as a direct input to different models of our systems, which generates expected results.

After that, the system uses the Parallel CNN and Transformer-Encoders [71] model taking Mel-spectrograms as features to classify the audio signal's seven emotional states of the user's speech, i.e., anger, happiness, disgust, sadness. The emotional state will help further models to get the context of the speech.

Then, the spectrograms of speech audios are fed into an RNN-based speech-to-text model DeepSpeech [22], to generate an English text transcript. The default DeepSpeech model does not produce expected transcription well for the southeast Asian accent for the English language. We get the word error rate (WER) of 0.44. That is why we tuned the DeepSpeech model on a consortium of a high-quality corpus of 13 major Indian languages [8], which achieved a WER of 0.18.

At this moment, our system knows the "emotional state" of the user's speech and a transcription of what the user says to the system. These two attributes will be used by a multi-domain conversational agent [18] to generate contextual reply text for the user. The reply texts of the conversational agents will further be used for text-to-speech models.

Furthermore, After getting reply texts from the conversational agent, the system will feed them into Flowtron [64] which will not only generate Mel-spectrograms for speech synthesis from the text but also control different aspects of speech synthesis such as pitch, tone, speech rate, and accent. This will make the synthesized speech as human likely as possible. Also, emotional states can be added with these synthesized voices by transferring styles of given data. For example, if we want to generate an angry state of the synthesized voice, we will give angry emotional audio clips to the trained model function, and it will manipulate the synthesized speech to generate an angry version. Mel-spectrograms generated by Flowtron will be used by another model called WaveGlow [51] to generate speech audio signals. Thus, the user will hear conversational agents' replies with a proper human-like voice along with human-like emotional states poured into the synthesized voice.

## 5.1 Model specification

### 5.1.1 2D parallel CNN and transformer-encoders

To take advantage of CNN's (Fig. 7) image classification and feature representation capacity, we need to represent our extracted audio features like MFCC, and the Mel Spectrogram graph as an image. Each value of the MFCC/Mel Spectrogram is the amplitude of the audio at a given Mel frequency range at a given time. Transformers (Fig. 8) are particularly good at predicting future frames/data. Since this is time-series information, we can use the Transformer to find the temporal relationship between pitch change and predict the future frequency distribution of particular emotions. This approach is the successor to the LSTM-RNN model that we tried earlier in our experiment. We use the Mel spectrogram as our experimental feature for this model. Like all previous ones, this classification has seven emotional groups and four emotional datasets. Not all data is distributed proportionally. We need to divide them into train, test, and validation data while preserving proportionality.

Utilizing the wisdom of previous CNN papers' findings, the proposed model was developed [71]. Conv, Pool, Conv, Pool, FC layer pattern was implemented in the architecture of LeNet. AlexNet presents the idea of increasing the sophistication of features by channel expansion using stacked CNNs. Parallelization was inspired by GoogLeNet [62] and Inception, which lets us diversify the features we learn from the data. The idea of using a smaller size kernel comes from VGGNet, which replaces AlexNets (11 x 11), stride 5 with (3 x 3) kernel, and gains significant improvement over it.

CNN with 2D Conv layers is the de facto methodology for image processing. For our case (Fig. 9), we have to imagine the Mel-Spectrogram plot as a single channel black and white image. There are two primary reasons for using two stacked filters: feature sophistication and efficiency. If we stack three layers of (3 x 3) kernels, in the second stack, the kernel will see a (5 x 5) view, and the third stack will see a (7 x 7) view of the original input. On the other hand, If we used a single (7 x 7) layer, it would have performed only a linear transformation. Moreover, we have been able to minimize excessive computation by using a stacked kernel. If we take the channel as constant, then for (3 x 3) kernel, we will have $27C^2$ parameters, whereas (7 x 7) kernel will have $49C^2$ parameters. In summary, using smaller stacked kernels, we are getting more intricate features and making the model more efficient. The sequential expansion of filter complexity and reduction in feature maps will give us the best hierarchical features with the lowest possible computation cost.

The motivation for the transformer encoder is to learn the temporal features and hope that it will be able to learn the frequency distribution of different emotions according to the global structure of the Mel-spectrogram of each emotion. RNN-LSTM was a possible candidate for this job, but it would have learned to predict the frequency changes according to time steps. The nature of the Transformer allows it to look at multiple different timestamps using a multi-head self-attention layer, which will, in turn, let us predict the next. As the transformers are very good at generating sequential data, the author expected it to perform well by looking at the entire sequence of frequencies, not just one timestamp. Max-Pooling the input Mel-Spectrogram map to the Transformer dramatically reduces the complexity and number of parameters.

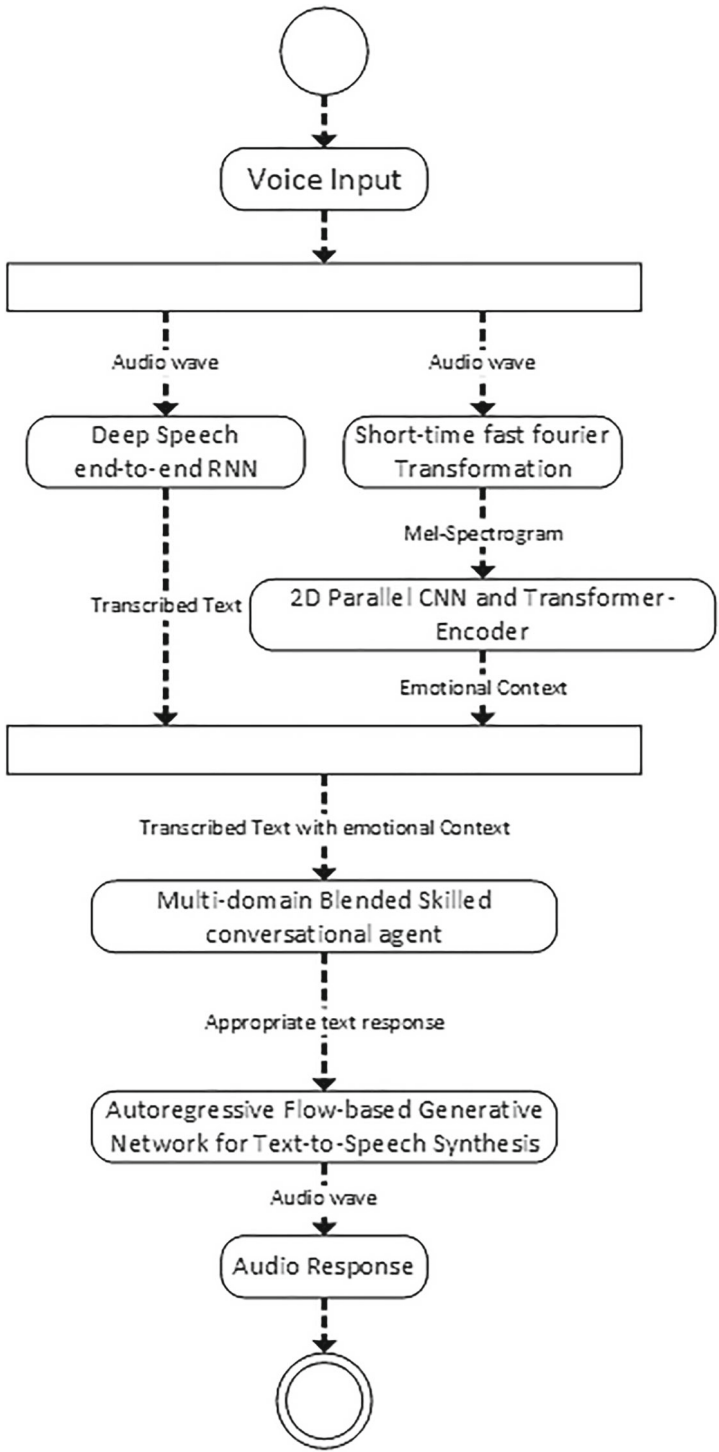

**Fig. 6** The block diagram of proposed affective social anthropomorphic intelligent system

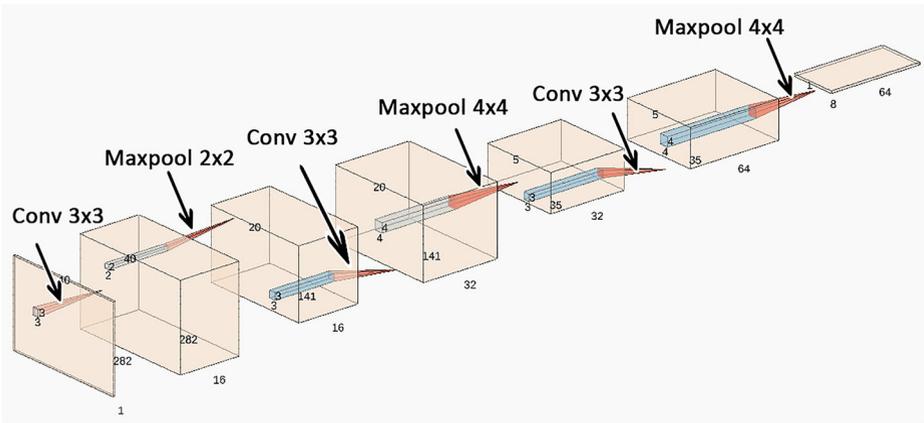

**Fig. 7** CNN architecture

Initially, the "Adam" optimizer was used because it usually works decently out of the box. But due to the fact that better performance is achievable by the good old SGD, the author changed the optimizer later with the highest momentum leading to convergence and acceptably long training time.

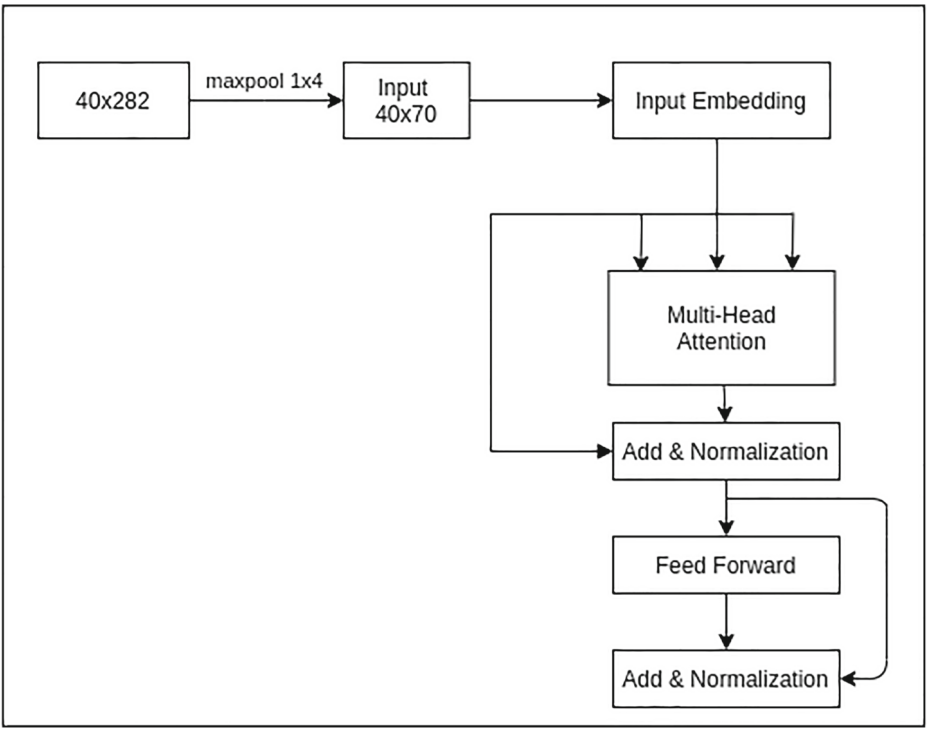

**Fig. 8** Transformer-encoder

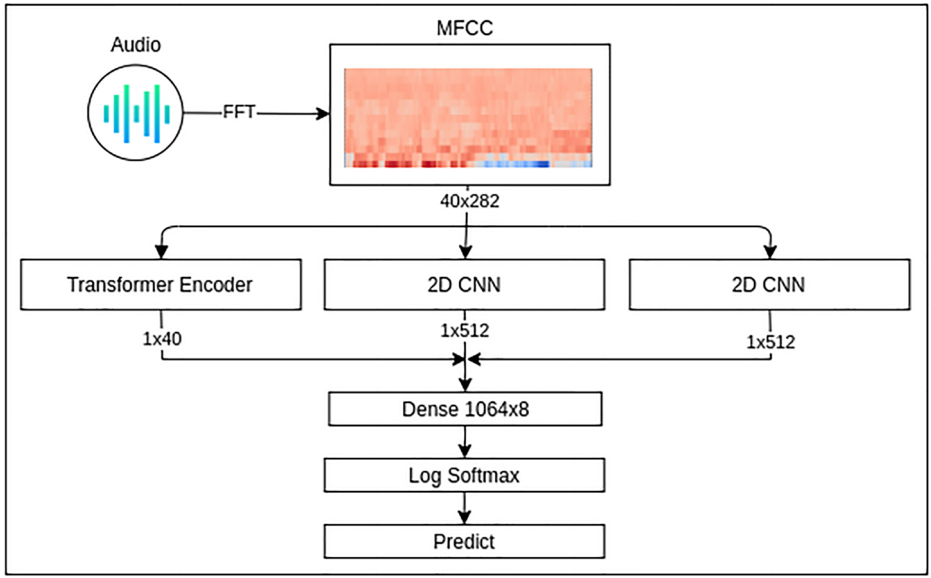

**Fig. 9** 2D prallel CNN-transformer-encoder

### 5.1.2 DeepSpeech: end-to-end RNN

At Silicon Valley AI Lab, Baidu researchers have made a well-optimized end-to-end RNN speech training system (Fig. 10) called "Deep Speech" with novel data synthesis techniques to obtain ample amounts of varied data for training achieving a 19.1% error rate on noisy speech dataset produced by them [22]. The system takes spectrograms of speech audios and generates the text transcription in English. The training set that is arranged for this system is, $X = \{(x^{(1)}, y^{(1)}), (x^{(2)}, y^{(2)}), ...\}$ where $x$ is a single utterance and $y$ is denoted as a label. A single utterance $x^{(i)}$ is a collective of vectors of audio features in a time-series of length $T^{(i)}, x_t^{(i)}; t = 1, 2, ..., T^{(i)}$. The objective of the RNN is to convert an input sequence x into a character probability sequence for the text transcription $\hat{y}_t = \mathbb{P}(c_t|x); c_t \in \{a, ...z, space, blank, apostrophe\}$ [22]. The RNN model comprises five hidden layers. The units of the hidden layer are denoted as $h^{(l)}; l$ represents each layer. Among the hidden layers, the first three layers are non-recurrent. The fourth layer is a bi-directional recurrent layer [56]. The fifth hidden layer unites both forward and backward units of bi-directional recurrent layers. At first, the first layer takes spectrogram frame $x_t; t = eachtimeslice$ as well as a context of $C$ frames. For each time step $t$, the second and third non-recurrent layers work by taking independent data. The computational function of the first three layers is:

$$h_t^{(l)} = g(W^{(l)}h^{(l-1)} + b^{(l)});\tag{1}$$

$$W^{(l)} = weight, b^{(l)} = bias, l = currentlayer\tag{2}$$

where $g(x) = min\{max\{0, x\}, 20\}$ is a rectified-liner (ReLu) activation [3] function. After that, the fourth bi-directional layer is created by two hidden units: forward recurrence $h^{(f)}$ and backward recurrence $h^{(b)}$. The computational function of both units is:

$$h_t^{(f)} = g(W^{(4)}h_t^{(3)} + W_r^{(f)}h_{t-1}^{(f)} + b^{(4)})\tag{3}$$

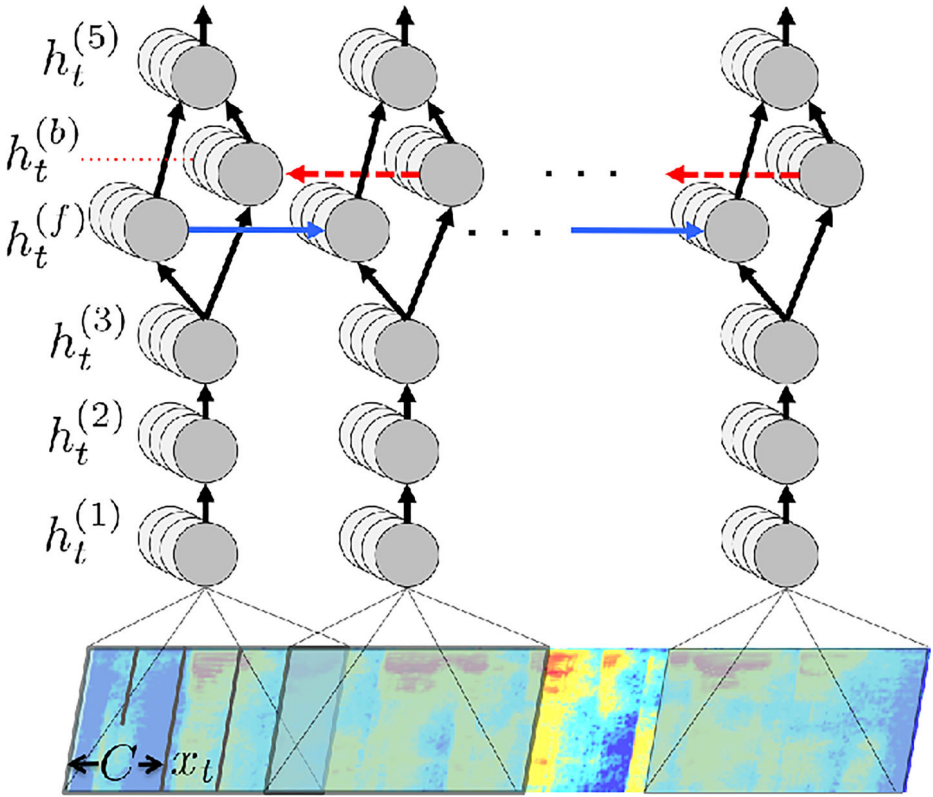

**Fig. 10** Structure of DeepSpeech's RNN model and notation

$$h_t^{(b)} = g(W^{(4)}h_t^{(3)} + W_r^{(b)}h_{t-1}^{(b)} + b^{(4)}) \tag{4}$$

In the case of forwarding recurrence, for each utterance $i$, $h^{(f)}$ is computed sequentially $t = 1$ to $t = T^{(i)}$. On the other hand, for backward recurrence $h^{(b)}$ is computed sequentially in reverse order, $t = T^{(i)}$ to $t = 1$. Both forward and backward hidden layers are combined and fed into the fifth layer. The computational function of the fifth layer, which is not recurrent, is:

$$g(W^{(5)}h_t^{(4)} + b^{(5)}); h_t^{(4)} = h_t^{(f)} + h_t^{(b)} \tag{5}$$

Finally, the output layer predicts the character probabilities with the help of the standard SoftMax function [22]:

$$h_{t,k}^{(6)} = \hat{y}_{t,k} \equiv \mathbb{P}(c_t = k|x) = \frac{\exp W_k^{(6)}h_t^{(5)} + b_k^{(6)}}{\sum_j \exp W_j^{(6)}h_t^{(5)} + b_j^{(6)}} \tag{6}$$

$$W^{(6)} = k^{th} column of weight matrix, b_k^{(6)} = k^{th} bias \tag{7}$$

### 5.1.3 Multi-domain conversational agent

To have a neutral conversation, an agent must have several skills, such as being engaging, knowledgeable, and empathetic, while sticking to the personality. Many prior approaches sought to acquire these abilities in isolation, but the actual human-like conversation goal was not accomplished. A team of Facebook researchers showed that these skills could be taught to a broad number of models if we provide adequate training data and generation strategy. Recently published, Blended Skill Talk [60] (BST) offers conversational context and training data that can be used to train multi-domain human-like conversational agents. The generation algorithm is also an indispensable part of the process. A model with the same accuracy but with a different generation algorithm can produce a completely different result. The authors also noted that the length of utterance plays a significant role in human judgment, whether it is engaging or not. According to the experiment, a too-short utterance can make the human judge perceive the bot as dull and uninterested. On the other hand, too-long utterances make the judge feel the bot is not listening and is distracted. Despite the previous report of beam searching being inferior to sampling [2, 24] the study shows that by tweaking the minimum beam length, control over the dull versus spicy response generation can be achieved. In this study, three types of architecture were used: Retriever, Generative, and retrieve-and-refine models. All of which were derived from the Transformer model (Fig. 11).

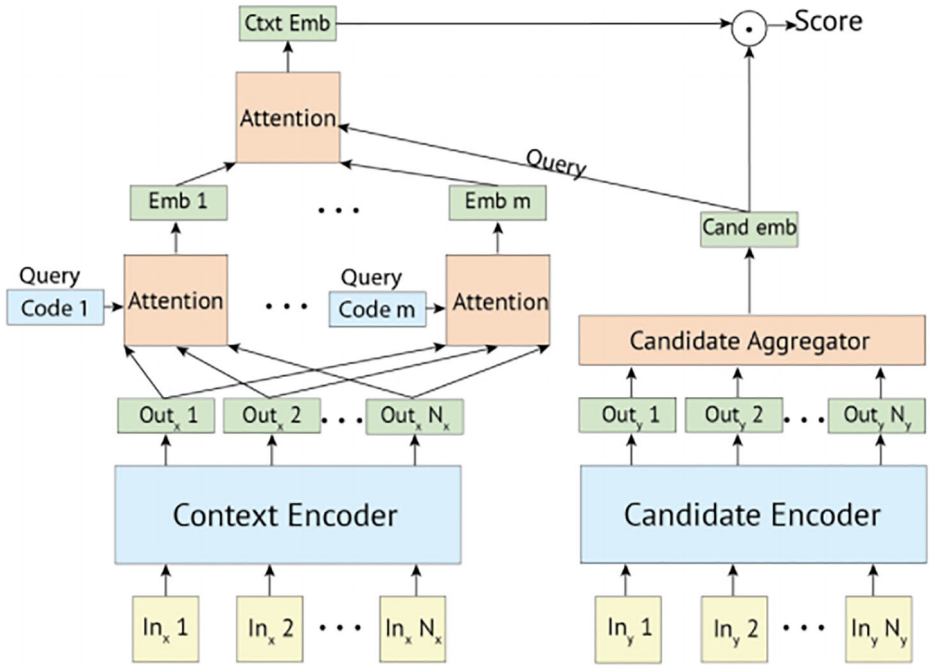

**Fig. 11** The Poly-Encoder Transformer architecture for retrieval

The Retriever model works by scoring the set of possible responses and outputting the highest probable one, given we have conversation history as input. The researchers used poly-encoder architecture [25] to encode global features of the context using several representations attended to by each potential candidate response [52]. The final attention mechanism allows us to achieve better performance over a single global vector representation. It generates context embedding and dot product it with each response candidate. The embeddings are created in two steps. Firstly, the model gets the candidate embedding using a transformer-based encoder and an aggregator function that takes the classifier embedding output or the token's mean. After that model encodes the context using another transformer and performs an "m" attention block. Each attention uses the Transformer output as keys and values, and the learned ci code is unique for each attention. On top of this embedding, another attention is calculated. The key and values are the output from the other attention.

$$Transformer output, T(x) = (h_1, ..., h_N) \tag{8}$$

$$y_{ctxt}^i = \sum_j w_j^{c_i} h_j \tag{9}$$

$$(w_1^{c_i}, ..., w_N^{c_i}) = softmax(c_i.h_1, ..., c_i.h_N) \tag{10}$$

$$y_{ctxt} = \sum_i w_i y_{xtxt}^i \tag{11}$$

$$(w_1, ..., w_m) = softmax(y_{cand_i}.y_c^1 txt, ..., y_{cand_i}.y_c^m txt \tag{12}$$

One of the most significant benefits of poly-encoders is that it gives a state-of-the-art performance on some dialogue tasks compared to other retrieval methods on ConvAI2 competition tasks based on human evaluation.

The generator approach is similar to the seq2seq model proposed in the Transformer [65] paper, but the main difference is that it is a lot bigger. For comparison, Google's Meena [2] has 2.7B parameters, whereas the blender model has 90M, 2.7B, 9.4B parameter versions.

Lastly, there is the retrieve and refine the approach. It mixes the previously mentioned two models. The retrieval model's output goes as an input of the generative model using a unique separator token. Utilizing this method, the authors tried to mitigate the known shortcomings like knowledge hallucination, disability to read new and external knowledge, dull and repetitive answers. They worked with two types of retrieval models: dialogue retriever and knowledge retriever. Dialogue retriever uses dialogue history to generate a response. Knowledge retriever gets its information from a large knowledge base. In this scenario, a transformer is trained to determine whether a knowledge retriever should be used (Fig. 12).

For decoding, Beam Search, Top-K-sampling, and sample-and-rank-sampling strategies were used. There are many different algorithms to decode the final output sequence as our model gives a probability distribution over the vocabulary. We need to select one word at a time until we reach the end of the statement. We can use greedy algorithms to choose the best word each time, but the final result may not be the overall best probable sentence. To mitigate this, we predict beam_size (possible sentences). At each step, we predict the next beam_size token for each sentence and select the one with the most probable beam_size. We stop if we reach the end character (complete "n" sentences) or after t steps. Next comes the Top-K sampling algorithm. Here, at each step, the word "i" is chosen by sampling the model distribution from the "k" most likely candidates. Along with the decoding process, some additional constraints were tested. Minimum length forced the model to produce a result with a defined length. Another one was a predictive retriever model, which predicts the sentence's length and limits the generation to that length. The last one was beam blocking, where the model was forced not to generate any trigram, a group of 3 words, in the next utterance if that is in the input or utterance itself.

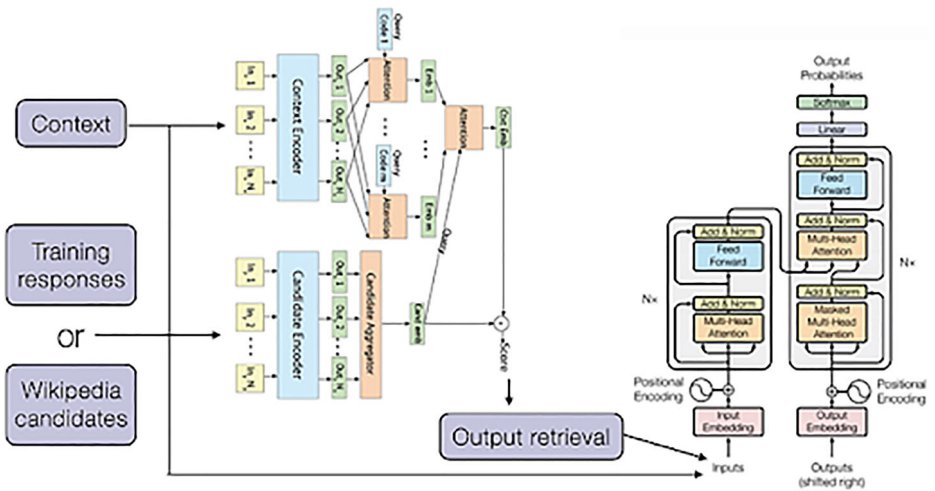

**Fig. 12** The retrieve and refine architecture

### 5.1.4 Autoregressive flow-based generative network for TTS synthesis

Text is required for Mel-spectrogram synthesis, which will have non-textual information such as tone, accent, and pitch. Also, non-textual information needs to follow the style of the given audio data. If we give the model some audio data of a particular emotion, such as anger, surprise, etc., the synthesized Mel-spectrogram should copy the style we refer to as "Style Transfer." NVIDIA researchers have introduced a model called "Flowtron," which does exactly the same thing as mentioned above. Flowtron does this by maximizing the probability of training data. Flowtron learns an invertible mapping of data to a latent space that can be manipulated to control many aspects of speech synthesis [64], including pitch, accent, speech rate, tone, etc. It generates a Mel-spectrogram frame based on previous Mel-spectrogram frames.

The whole sequence of frames is $p(X) = \prod p(x_t|x_{1:t-1})$. Two types of distributions, $p(z)$, are used to be sampled by the neural network, which is used as a generative model in the flowtron. The first distribution is a zero-mean spherical Gaussian, $z \sim N(z; 0, I)$. The other one is a mixture of spherical Gaussian with fixed or learnable parameters.

$$z \sim \sum_k \hat{\phi}_k N\left(z; \hat{\mu}_k, \hat{\sum}_k\right) \tag{13}$$

The samples are transformed into $p(x)$ from $p(z)$ by going through "affine transformations," which are invertible and parameterized transformation. We know that flowtron uses an invertible neural network. Invertible neural networks are constructed using coupling layers [28, 29], in this case, affine coupling layer [16]. For each input $x_{t-1}$ a scale, s , and a bias is produced. This scale and bias affine transforms the next input $x_t$:

$$(logs_t, b_t) = NN(x_{1:t-1}, text, speaker) \tag{14}$$

$$x'_t = s_t \odot x_t + b_t \tag{15}$$

Here, $NN()$ denotes any autoregressive causal transformation. A zero vector is concatenated with other inputs of $NN()$ to implement this. The $NN()$ needs not to be invertible, but the affine coupling layer preserves the whole network's invertibility. In the autoregressive structure, every t-th variable $z'_t$ depends on its previous timesteps from the star $z_{1:t-1}$ : [30].

$$z'_t = f_k(z_{1:t-1}) \tag{16}$$

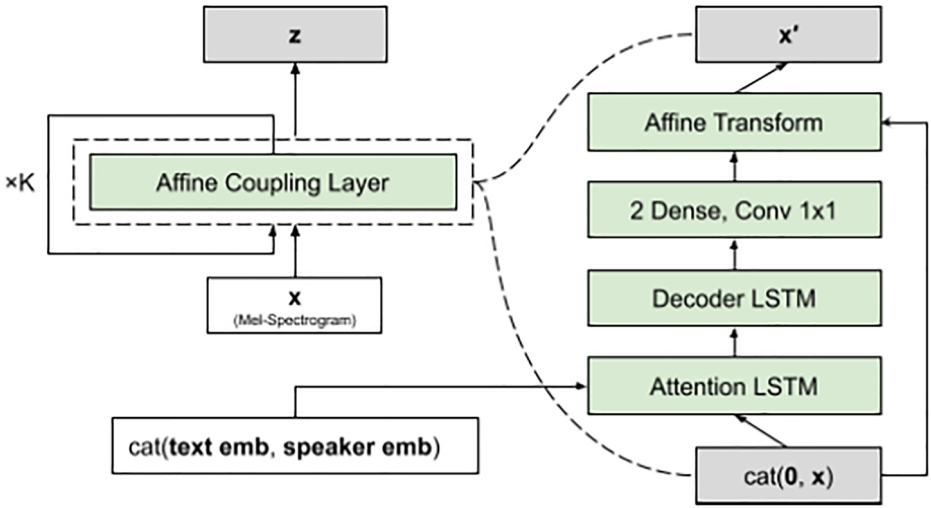

**Fig. 13** Flowtron Network. Text and speaker embeddings are channel-wise concatenated. A 0-values vector is concatenated with x in the time dimension

Flowtron is maximizing the data's log-likelihood by utilizing the parameterized affine transformation and the autoregressive structure mentioned above. These are possible by using the change of variables:

$$\log p_\theta(x) = \log p_\theta(z) + \sum_{i=1}^{k} \log |\det(j(f_i^{-1}(x)))| \qquad (17)$$

$$z = f_k^{-1} \circ f_{k-1}^{-1} \circ ... f_0^{-1}(x) \qquad (18)$$

Mel-spectrograms are converted as vectors and run through several affine coupling layers conditioned on the text and fixed dummy speaker embedding. Each affine coupling layer is called the "flow." Finally, the processed vectors are forwarded to pass through the neural network. Randomly sampled z values from Gaussian Mixture or spherical Gaussian with fixed or flowtron predicted parameters are run through the trained neural network to infer. The inferred Mel-spectrograms are decoded into waveforms using a single pre-trained WaveGlow [51] model trained on a single speaker (Fig. 13).

# 6 Performance evaluation

## 6.1 Performance measurements metrics

### 6.1.1 Word Error Rate (WER)

The word error rate is based on the Levenshtein distance [50]. It is computed by the minimum number of operations, i.e., insertion, deletion, substitution, to be performed to generate a text hypothesis that is similar to the reference text data. The computational function of WER is:

$$WER = \frac{1}{N^*_{ref}} \sum_{k=1}^{K} \min_r d_L(ref_{k,r}, hyp_k) \tag{19}$$

here $d_L(ref_{k,r}, hyp_k)$ is the Levenshtein distance from $hyp_k$ to $ref_{k,r}$.

### 6.1.2 F1- score

F1 score is the weighted average of Precision and Recall. Therefore, this score takes both false positives and false negatives into account. It is used to balance precision and recall. F1 score is better than accuracy in cases where class distribution is uneven. So, for our case, we took the F1 score as our measurement metric.

### 6.1.3 Mean opinion score

Mean opinion score (MOS) is a measure used in the video, audio, and audiovisual, representing the overall quality of a stimulus or system. It is the arithmetic mean over all individual "values on a predefined scale that a subject assigns to his opinion of the performance of a system quality." Such ratings are usually gathered in a subjective quality evaluation test, but they can also be algorithmically estimated. It is expressed as a single rational number, typically in the range 1–5, where 1 is the lowest perceived quality, and 5 is the highest perceived quality. This metric is calculated using the arithmetic mean over a single rating performed by humans.

$$MOS = \frac{\sum_{n=1}^{N} R_n}{N} \tag{20}$$

Here, R is the rating given for the clip, and N is the number of participants. We will compare two of our models using this metric to determine which one is better and also provide a real demonstration.

## 6.2 Model implementation

### 6.2.1 Sequential 1D CNN

We started our emotion classifier from scratch. To get the feel of the MFCC feature, we used the mean value of the feature to determine the class of the emotion using a 1D Convolutional Neural Network. The approach was too naive and was not able to produce a good result. We removed the gender class (reducing it to 7 from 14) to make it more predictable, but the best result we could produce is 50.28% accuracy with 100 Epoch, RMSprop as an optimizer, and MFCC value of 13 (Fig. 14, Tables 1 and 2).

After seeing the result, we came to the conclusion that it would not be a very smart idea to spend on this approach, so we moved onto the next method.

### 6.2.2 Sequential 2D CNN

Then we used the MFCC values to create an image and use Convolutional Neural Network to classify the image. By classifying the image, we were able to classify the emotion as well. The first trial gave us a somewhat hopeful result, so we went further with it. We tried different parameters and tried to tweak the model. The initial accuracy was 67.08%. Using only the augmented data seemed to reduce the accuracy even more. When we added augmented data and real data together and trained the model with it, we got the best result. MFCC coefficient also plays a decent role in increasing accuracy as it increases the resolution of the feature. The number of epochs also positively influences the accuracy result to a certain degree. After that, we get diminishing returns. For 100 epochs, we got an accuracy of 72.26%. After that, we increased the epochs value by 50%, making it 150, but the result accuracy was increased by 0.01% (Fig. 15 and Table 3).

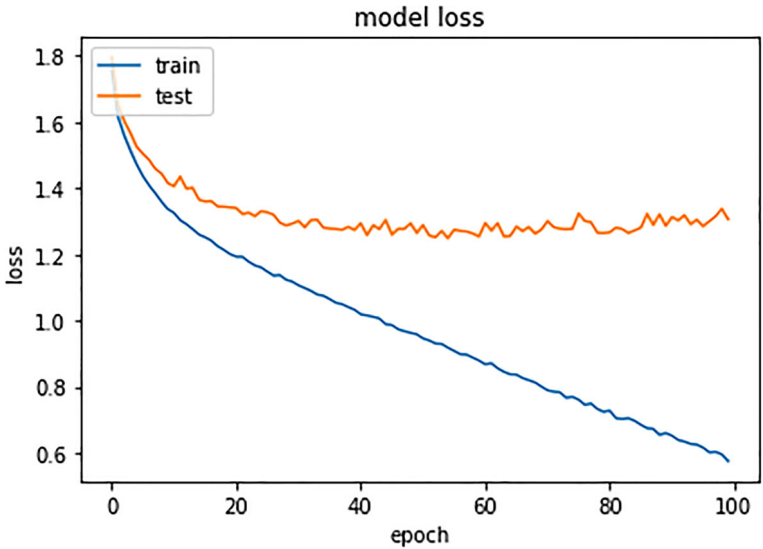

**Fig. 14** Best 1D-CNN loss graph

### 6.2.3 CNN-LSTM

We also tried the CNN-LSTM model. We only ran this model on a RAVDESS dataset with log mel spectrogram as a feature, but the result it produced was indeed promising, but from model loss graphs, we can get the idea that this model is not stable. The first time we ran the model, we got an accuracy of 59.02%, but after running the model again, the accuracy jumped to 72.91%. We ran the experiment several times but could not make the model stable enough for our use. Thus, we abandoned the method (Fig. 16 and Table 4).

**Table 1** Best 1D-CNN F1 score matrix

|  | Precision | Recall | F1-Score | Support |
|---|---|---|---|---|
| Angry | 0.61 | 0.66 | 0.63 | 489 |
| Disgust | 0.41 | 0.52 | 0.46 | 478 |
| Fear | 0.41 | 0.37 | 0.39 | 460 |
| Happy | 0.56 | 0.41 | 0.47 | 498 |
| Neutral | 0.42 | 0.54 | 0.47 | 453 |
| Sad | 0.54 | 0.47 | 0.50 | 496 |
| Surprised | 0.89 | 0.65 | 0.75 | 167 |
| Accuracy | — |  | 0.50 | 3041 |
| Macro avg | 0.55 | 0.52 | 0.53 | 3041 |
| Weighted avg | 0.52 | 0.50 | 0.50 | 3041 |

**Table 2** Comparison of 1D models

| Epoch | Class | Accuracy | Augmentation |
|-------|-------|----------|--------------|
| 100 | 14 | 42.98% | no |
| 100 | 14 | 49.02% | no |
| 100 | 7 | 50.28% | no |

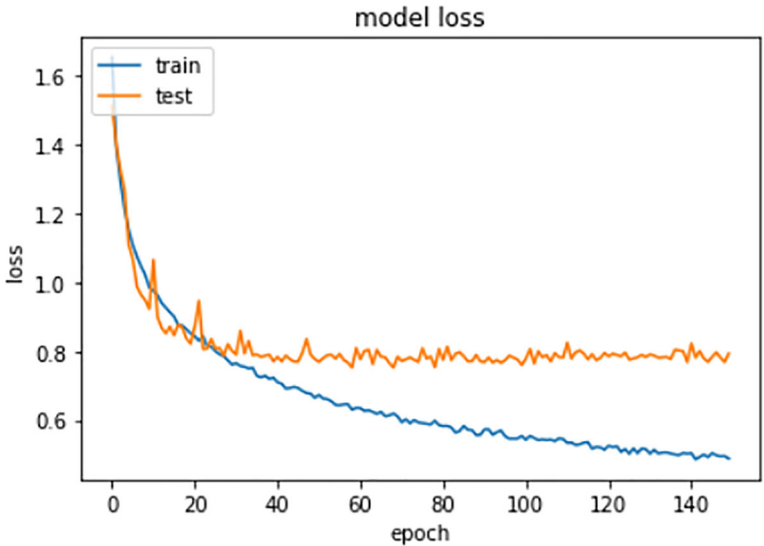

**Fig. 15** Best 2D-CNN loss graph

**Table 3** Comparison of 2D CNN models

| Epoch | Class | Accuracy | Parameters | Augmentation |
|-------|-------|----------|------------|--------------|
| 100 | 7 | 67.08% | mean(n_mfcc =30) | no |
| 50 | 7 | 67.58% | mean(n_mfcc =30) | no |
| 35 | 7 | 69.58% | mean(n_mfcc =30) | no |
| 100 | 7 | 69.42% | mean(n_mfcc =30) | yes |
| 50 | 7 | 66.95% | mean(n_mfcc =30) | yes |
| 35 | 7 | 68.07% | mean(n_mfcc =30) | yes |
| 100 | 7 | 68.33% | mean(n_mfcc =50) | no |
| 50 | 7 | 68.79% | mean(n_mfcc =50) | no |
| 35 | 7 | 69.45% | mean(n_mfcc =50) | no |
| 100 | 7 | 72.26% | mean(n_mfcc =30) | yes |
| 50 | 7 | 70.84% | mean(n_mfcc =30) | yes |
| 35 | 7 | 69.79% | mean(n_mfcc =30) | yes |
| 150 | 7 | 72.27% | mean(n_mfcc =30) | yes |

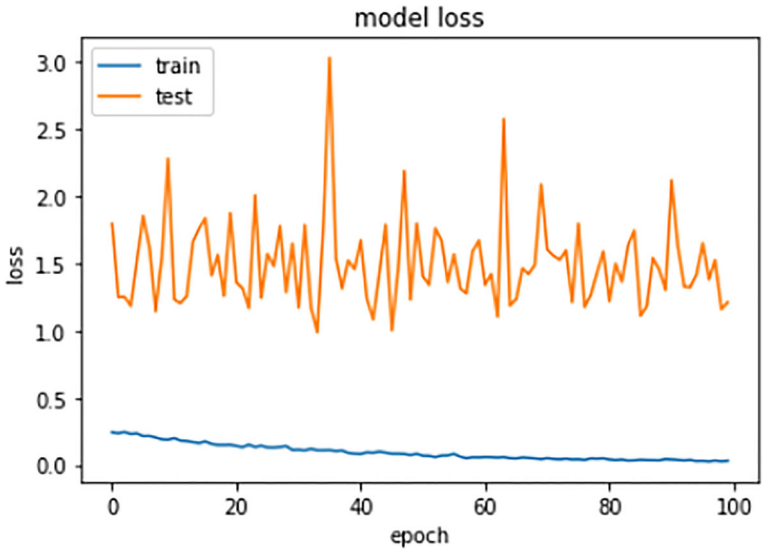

**Fig. 16** CNN-LSTM loss graph

### 6.2.4 XResnet models (transfer learning)

xResnet50 [23] is one of the most popular architectures in computer vision research. We tried to experiment with xResnet50 and xResnet18, which are relatively small in size but good in terms of training speed and accuracy. We used both heavy and light augmentation for the experiment and observed the performance. The augmentation includes removing silence, the addition of white noise, signal loss, changing volume, resizing and using different spectrogram-augmentation [46], which modifies the spectrogram to gain robustness against the deformation of spectrograms in the time direction. For features, we have chosen MFCC images with delta, and Mel-Spectrogram with parameters optimized for voice speeches to experiment with. We did not get any significant accuracy improvement from these experiments. We also found that applying heavy augmentation decreases the accuracy of the models (Table 5).

**Table 4** CNN-LSTM accuracy result

| Epoch | Optimizer | Class | Accuracy | Augmentation |
|-------|-----------|-------|----------|--------------|
| 100 | SGD | 8 | 72.91% | yes |

**Table 5** Accuracy comparison of various xResNet models

| Model | Feature | Accuracy | Epoch | Augmentation |
|---|---|---|---|---|
| xresnet50 | MFCC + Delta | 0.6986 | 25 | RemoveSilence, AddNoise, SignalLoss, ChangeVolume, resize |
| xresnet18 | MFCC + Delta | 0.7018 | 25 | AddNoise, resize |
| xresnet50 | Mel Voice | 0.7150 | 15 | AddNoise, resize |
| xresnet50 | Mel Voice | 0.6809 | 15 | RemoveSilence(Trim), MaskTime(size=4), MaskFreq(size=10), Resize |

### 6.2.5 VGG19 (transfer learning)

VGG19 [59] is another version of the VGG model with 19 layers, including 16 convolution layers, three fully connected layers along with five max-pooling layers, and one SoftMax layer. This model is a relatively older and simple model, but still, it is an effective one. That motivated us to experiment with this model. The reason behind our experimentation with VGG19 is that it is just another decent classification architecture for images that works well despite being very simple and transfer learning is possible with this architecture. Just like xresnet models, we used MFCC and delta features for VGG19. We also augmented the datasets by removing silence, resizing the images, changing volume, adding noise, and signal loss. However, the model did not perform up to the mark for the datasets that we used (Table 6).

### 6.2.6 Parallel 2D CNN with transformer-encoder

Parallel 2D CNN has a sequential expansion of filter complexity. It reduces feature maps gradually, which gives us high-quality features with better computational performance. Moreover, the transformer encoder learns the frequency distribution of different emotional categories by focusing on the temporal features. Mel-spectrograms of the audio signals have been used as a feature for this model. We have added Gaussian white noise to the data as a procedure of augmentation. We have got noticeable accuracy for the respective dataset that we used for this model (Table 7).

### 6.2.7 DeepSpeech tuning

We have used the DeepSpeech [22] model for the generation of English transcripts from speech audio clips. However, the baseline model did not come up with a promising output for input data given by us. We assumed that it might happen for our accent as we are from the southeast Asian region and not native English speakers. That's why we tuned the DeepSpeech model with the Indic TTS [27] project. We had to clean the articles from the datasets and prepared them in a structure that is suitable for the deep speech model. As per our assumption, the tuned model gave a much better result. Better accuracy is possible if we train it with more datasets and for a longer time (Table 8).

**Table 6** Accuracy comparison of VGG19 models

| Feature | Accuracy | Epoch | Augmentation |
|---|---|---|---|
| MFCC + Delta | 0.707237 | 15 | RemoveSilence(Trim), AddNoise, SignalLoss, ChangeVolume, resize |
| MFCC + Delta | 0.685033 | 15 | AddNoise, resize |

**Table 7** Accuracy of Parallel 2D CNN with Transformer Encoder model

| Feature | Accuracy | Epoch | Augmentation |
| --- | --- | --- | --- |
| Mel Spectrogram | 0.8667 | 400 | AddGausianWhiteNoise |

## 7 Conclusion & future work

The proposed anthropomorphic intelligence system yielded promising results from the chained models on which we performed our research. We have tried several techniques to bridge the human-IVA communication gap. Several hurdles have arisen throughout the model's implementation. NLP is a rapidly growing field; keeping up with the latest and best technologies is quite difficult. We sought to include the most outstanding and most current innovations in these domains to get the greatest possible outcome for our system. Various style data were added to Flowtron's pre-trained models to tune further and enhance our models. The open-domain conversational model performed as expected. We were also quite pleased with the accuracy of the Speech-to-Text model that we customized for our region. Overall, the system we built functioned far better than we anticipated. Unfortunately, there are still many undiscovered avenues. We observed that there are no publicly available emotion style datasets. We want to generate an emotional style dataset that will be used for audio style transfer. We also believe that the Parallel 2D CNN Transformer-Encoder model can be improved by tweaking its 2D CNN models. We want to explore transfer learning and the benefits of reducing negative transfer in speech emotion recognition. In the future, we also want to be able to recognize and eliminate numerous background voices in the input environment. Similarly, while the user's audio is being delivered, countless other auditory distractions, such as music, may be present in the background. In these instances, we would want to extract audio data from the environment utilizing some of the approaches described in the literature review, such as pitch-based audio data separation from music and characteristic-based detection [7].

**Table 8** DeepSpeech accuracy improvement

| Metric | Trained Model | DeepSpeech Model |
| --- | --- | --- |
| Word Error Rate | 0.18 | 0.44 |

**Acknowledgements** The authors are grateful to King Saud University, Riyadh, Saudi Arabia for funding this work through Researchers Supporting Project Number-RSP2023R18.

**Funding** Open access funding provided by SINTEF.

**Data Availability** This datasets used in this work were already publicly available. Datasets used for Speech Emotion Recognition. RAVDESS: https://www.kaggle.com/datasets/uwrfkaggler/ravdess-emotional-speech-audio, SAVEE: https://www.kaggle.com/datasets/barelydedicated/savee-database, TESS: https://www.kaggle.com/datasets/ejlok1/toronto-emotional-speech-set-tess, CREMA-D: https://www.kaggle.com/datasets/ejlok1/cremad. Dataset used for Speech-to-Text. Indic TTS: https://bhaasha.iiit.ac.in/indic-tts/. Dataset used for text-to-Speech + style transfer. LJ speech: https://keithito.com/LJ-Speech-Dataset/, Libri TTS: https://paperswithcode.com/dataset/librispeech.

## Declarations

**Conflict of Interests** The authors have no conflicts of interest regarding this work.

## References

1. Aazam B, Dariush A, Mahdi H (2019) Increasing the accuracy of automatic speaker age estimation by using multiple UBMS. In: 2019 5th conference on knowledge based engineering and innovation, KBEI, pp 592–598
2. Adiwardana D, Luong M-T, So DR, Hall J, Fiedel N, Thoppilan R, Yang Z, Kulshreshtha A, Nemade G, Lu Y, Le QV (2020) Towards a human-like open-domain chatbot. arXiv:2001.09977
3. Agarap AF (2019) Deep learning using rectified linear units (relu). arXiv:1803.08375
4. Akuzawa K, Iwasawa Y, Matsuo Y (2019) Expressive speech synthesis via modeling expressions with variational autoencoder. arXiv:1804.02135
5. Anagnostopoulos C-N, Iliou T, Giannoukos I (2015) Features and classifiers for emotion recognition from speech: a survey from 2000 to 2011. Artif Intell Rev 43(2):155–177, 2
6. Aouam D, Benbelkacem S, Zenati N, Zakaria S, Meftah Z (2018) Voice-based augmented reality interactive system for car's components assembly. In: 2018 3rd international conference on pattern analysis and intelligent systems, PAIS, pp 1–5
7. Bastanfard A, Amirkhani D, Naderi S (2020) A singing voice separation method from persian music based on pitch detection methods. In: 2020 6th iranian conference on signal processing and intelligent systems, ICSPIS, pp 1–7
8. Baby A, Thomas AL Resources for indian languages, p 8
9. Chang J, Scherer S (2017) Learning representations of emotional speech with deep convolutional generative adversarial networks. arXiv:1705.02394
10. Cheyneycomputerscience/crema-d Crowd sourced emotional multimodal actors dataset (crema-d). Accessed 06 Oct 2020. https://github.com/CheyneyComputerScience/CREMA-D
11. Creswell A, White T, Dumoulin V, Arulkumaran K, Sengupta B, Bharath AA (2018) Generative adversarial networks: an overview. IEEE Signal Process Mag 35(1):53–65, 1
12. Dellaert F, Polzin T, Waibel A (1996) Recognizing emotion in speech. In: Proceeding of fourth international conference on spoken language processing. ICSLP '96, vol. 3. IEEE, Philadelphia, pp 1970–1973. Accessed: 01 Apr 2020. http://ieeexplore.ieee.org/document/608022/
13. Deng J, Xia R, Zhang Z, Liu Y, Schuller B (2014) Introducing shared-hidden-layer autoencoders for transfer learning and their application in acoustic emotion recognition. In: 2014 IEEE international conference on acoustics, speech and signal processing (ICASSP), pp 4818–4822

14. Deng J, Xu X, Zhang Z, Fruhholz S, Schuller B, Deng J, Xu X, Zhang Z, Fruhholz S, Schuller B (2018) Semisupervised autoencoders for speech emotion recognition. IEEE/ACM Trans Audio Speech Lang Proc (TASLP) 26(1):31–43, 1

15. Devillers L, Vidrascu L, Lamel L (2005) Challenges in real-life emotion annotation and machine learning based detection. Neural Netw 18(4):407–422, 5

16. Dinh L, Sohl-Dickstein J, Bengio S (2017) Density estimation using real NVP. arXiv:1605.08803

17. El Ayadi M, Kamel MS, Karray F (2011) Survey on speech emotion recognition: features, classification schemes, and databases. Pattern Recog 44(3):572–587, 3

18. Fierro C (2020) Recipes for building an open-domain chatbot. Accessed: 31 Dec 2020. https://medium.com/dair-ai/recipes-for-building-an-open-domain-chatbot-488e98f658a7

19. Gao Y, Singh R, Raj B (2018) Voice impersonation using generative adversarial networks. arXiv:1802.06840

20. Glorot X, Bordes A, Bengio Y Domain adaptation for large-scale sentiment classification: a deep learning approach. p 8

21. Goodfellow IJ, Pouget-Abadie J, Mirza M, Xu B, Warde-Farley D, Ozair S, Courville A, Bengio Y (2014) Generative adversarial networks. arXiv:1406.2661

22. Hannun A, Case C, Casper J, Catanzaro B, Diamos G, Elsen E, Prenger R, Satheesh S, Sengupta S, Coates A, Ng AY (2014) Deep speech: Scaling up end-to-end speech recognition. arXiv:1412.5567

23. He T, Zhang Z, Zhang H, Zhang Z, Xie J, Li M (2018) Bag of tricks for image classification with convolutional neural networks. arXiv:1812.01187

24. Holtzman A, Buys J, Du L, Forbes M, Choi Y (2020) The curious case of neural text degeneration. arXiv:1904.09751

25. Humeau S, Shuster K, Lachaux M-A, Weston J (2020) Poly-encoders: transformer architectures and pre-training strategies for fast and accurate multi-sentence scoring. arXiv:1905.01969

26. Ito K, Johnson L (2017) The lj speech dataset. Available: https://keithito.com/LJ-Speech-Dataset/

27. Indic tts. Accessed 07 Jan 2021. https://www.iitm.ac.in/donlab/tts/index.php

28. Kingma DP, Ba J (2017) Adam: a method for stochastic optimization. arXiv:1412.6980

29. Kingma DP, Dhariwal P (2018) Glow: generative flow with invertible 1x1 convolutions. arXiv:1807.03039

30. Kingma DP, Salimans T, Jozefowicz R, Chen X, Sutskever I, Welling M (2017) Improving variational inference with inverse autoregressive flow. arXiv:1606.04934

31. Kumar K, Kumar R, de Boissiere T, Gestin L, Teoh WZ, Sotelo J, de Brebisson A, Bengio Y, Courville A (2019) Melgan: Generative adversarial networks for conditional waveform synthesis. arXiv:1910.06711

32. Lee Y, Rabiee A, Lee S-Y (2017) Emotional end-to-end neural speech synthesizer. arXiv:1711.05447

33. Li J, Galley M, Brockett C, Spithourakis GP, Gao J, Dolan B (2016) A persona based neural conversation model. arXiv:1603.06155

34. Liu J, Chen C, Bu J, You M, Tao J (2007) Speech emotion recognition using an enhanced co-training algorithm. In: Multimedia and Expo, 2007 IEEE international conference on. IEEE, Beijing, pp 999–1002. Accessed: 02 Apr 2020. http://ieeexplore.ieee.org/document/4284821/

35. Livingstone SR, Russo FA (2018) The ryerson audio-visual database of emotional speech and song (ravdess): A dynamic, multimodal set of facial and vocal expressions in north american english. PLoS ONE 13(5):e0196391. publisher: Public Library of Science

36. Mohammad S, Aazam B (2013) Text material design for fuzzy emotional speech corpus based on persian semantic and structure. In: 2013 international conference on fuzzy theory and its applications (iFUZZY), pp 380–384

37. Mohammad S, Aazam B (2014) Study on unit-selection and statistical parametric speech synthesis techniques. J Comput Robot 7(1):19–25

38. Mohammad S, Azam B (2016) Real-time speech emotion recognition by minimum number of features. In: 2016 Artificial Intelligence and Robotics (IRANOPEN), pp 72–76

39. Montoya RM, Horton RS, Kirchner J (2008) Is actual similarity necessary for attraction? a meta-analysis of actual and perceived similarity. J Soc Pers Relationsh 25(6):889–922, 12

40. Morotti E, Stacchio L, Donatiello L, Roccetti M, Tarabelli J, Marfia G (2021) Exploiting fashion x-commerce through the empowerment of voice in the fashion virtual reality arena. Virtual Reality

41. Neff M, Wang Y, Abbott R, Walker M (2010) Evaluating the effect of gesture and language on personality perception in conversational agents. In: Allbeck J, Badler N, Bickmore T, Pelachaud C, Safonova A (eds) Intelligent Virtual Agents, vol. 6356, series Title: Lecture Notes in Computer Science. Springer, Berlin, pp 222–235, https://doi.org/10.1007/978-3-642-15892-6_24

42. Nishimura M, Hashimoto K, Oura K, Nankaku Y, Tokuda K (2016) Singing voice synthesis based on deep neural networks. In: Interspeech 2016, pp 2478–2482. Accessed 06 Oct 2020. http://www.isca-speech.org/archive/Interspeech_2016/abstracts/1027.html

43. Oord vdA, Dieleman S, Zen H, Simonyan K, Vinyals O, Graves A, Kalchbrenner N, Senior A, Kavukcuoglu K (2016) Wavenet: a generative model for raw audio. arXiv:1609.03499

44. Pan SJ, Yang Q (2010) A survey on transfer learning. IEEE Trans Knowl Data Eng 22(10):1345–1359, 10

45. Panayotov V, Chen G, Povey D, Khudanpur S (2015) Librispeech: an asr corpus based on public domain audio books. In: ICASSP 2015 - 2015 IEEE international conference on acoustics, speech and signal processing (ICASSP). IEEE, Queensland, pp 5206–5210. Accessed 07 Jan 2021. http://ieeexplore.ieee.org/document/7178964/

46. Park DS, Chan W, Zhang Y, Chiu C-C, Zoph B, Cubuk ED, Le QV (2019) Specaugment: a simple data augmentation method for automatic speech recognition. In: Interspeech 2019, pp 2613–2617. arXiv:1904.08779

47. Pascual S, Bonafonte A, Serrà J (2017) Segan: speech enhancement generative adversarial network. arXiv:1703.09452

48. Perez Garcia DM, Saffon Lopez S, Donis H (2018) Everybody is talking about virtual assistants, but how are people really using them? In: Proceedings of the 32nd international BCS human computer interaction conference. Accessed: 04 Apr 2020. https://scienceopen.com/document?vid=cebfbdb7-b28e-47ab-8b6e-c19015e12ab7

49. Pichora-Fuller MK, Dupuis K (2020) Toronto emotional speech set (tess). https://doi.org10.5683/SP2/E8H2MF. https://dataverse.scholarsportal.info/citation?persistentId=doi:10.5683/SP2/E8H2MF

50. Popović M, Ney H (2007) Word error rates: decomposition over pos classes and applications for error analysis, the Second Workshop. Prague, Czech Republic: Association for Computational Linguistics, pp 48–55. Accessed 08 Jan 2021. http://portal.acm.org/citation.cfm?doid=1626355.1626362

51. Prenger R, Valle R, Catanzaro B (2018) Waveglow: a flow-based generative network for speech synthesis. arXiv:1811.00002

52. Roller S, Dinan E, Goyal N, Ju D, Williamson M, Liu Y, Xu J, Ott M, Shuster K, Smith EM, Boureau Y-L, Weston J (2020) Recipes for building an open-domain chatbot. arXiv:2004.13637

53. Savargiv M, Bastanfard A (2015) Persian speech emotion recognition. In: 2015 7th conference on information and knowledge technology (IKT), pp 1–5

54. Schuller BW (2018) Speech emotion recognition: two decades in a nutshell, benchmarks, and ongoing trends. Commun ACM 61(5):90–99, 4

55. Schuller B, Batliner A (2014) Computational paralinguistics: emotion, affect and personality in speech and language processing, 1st. Wiley, Hoboken

56. Schuster M, Paliwal K (1997) Bidirectional recurrent neural networks. IEEE Trans Signal Process 45(11):2673–2681

57. Seyyed HM, Amir BA, Reza KM (2021) Trcla: a transfer learning approach to reduce negative transfer for cellular learning automata. In: IEEE transactions on neural networks and learning systems, pp 1–10

58. Shen J, Pang R, Weiss RJ, Schuster M, Jaitly N, Yang Z, Chen Z, Zhang Y, Wang Y, Skerry-Ryan RJ, Saurous RA, Agiomyrgiannakis Y, Wu Y (2018) Natural tts synthesis by conditioning wavenet on mel spectrogram predictions. arXiv:1712.05884

59. Simonyan K, Zisserman A (2015) Very deep convolutional networks for large-scale image recognition. arXiv:1409.1556

60. Smith EM, Williamson M, Shuster K, Weston J, Boureau Y-L (2020) Can you put it all together: Evaluating conversational agents' ability to blend skills. arXiv:2004.08449

61. Surrey audio-visual expressed emotion (savee) database. Accessed: 06 Oct 2020. http://kahlan.eps.surrey.ac.uk/savee/

62. Szegedy C, Liu W, Jia Y, Sermanet P, Reed S, Anguelov D, Erhan D, Vanhoucke V, Rabinovich A (2014) Going deeper with convolutions. arXiv:1409.4842

63. Trigeorgis G, Ringeval F, Brueckner R, Marchi E, Nicolaou MA, Schuller B, Zafeiriou S (2016) Adieu features? end-to-end speech emotion recognition using a deep convolutional recurrent network. In: 2016 IEEE international conference on acoustics, speech and signal processing (ICASSP). IEEE, Shanghai, pp 5200–5204. Accessed: 02 Apr 2020. http://ieeexplore.ieee.org/document/7472669/

64. Valle R, Shih K, Prenger R, Catanzaro B (2020) Flowtron: an autoregressive flow-based generative network for text-to-speech synthesis. arXiv:2005.05957

65. Vaswani A, Shazeer N, Parmar N, Uszkoreit J, Jones L, Gomez AN, Kaiser L, Polosukhin I (2017) Attention is all you need. arXiv:1706.03762

66. Wang K, Gou C, Duan Y, Lin Y, Zheng X, Wang F-Y (2017) Generative adversarial networks: introduction and outlook. IEEE/CAA J Autom Sin 4(4):588–598

67. Wang Y, Skerry-Ryan RJ, Stanton D, Wu Y, Weiss RJ, Jaitly N, Yang Z, Xiao Y, Chen Z, Bengio S, Le Q, Agiomyrgiannakis Y, Clark R, Saurous RA (2017) Tacotron: towards end-to-end speech synthesis. arXiv:1703.10135

68. Yamamoto R, Song E, Kim J-M (2020) Parallel wavegan: a fast waveform generation model based on generative adversarial networks with multi-resolution spectrogram. arXiv:1910.11480

69. You Q, Luo J, Jin H, Yang J Robust image sentiment analysis using progressively trained and domain transferred deep networks. arXiv:1509.06041

70. Zen H, Dang V, Clark R, Zhang Y, Weiss RJ, Jia Y, Chen Z, Wu Y (2019) Libritts: a corpus derived from librispeech for text-to-speech. arXiv:1904.02882

71. Zenkov I (2020) Transformer-cnn-emotion-recognition. GitHub, container-title: GitHub repository. Available: https://github.com/IliaZenkov/transformer-cnn-emotion-recognition

## Affiliations

**Md. Adyelullahil Mamun[1] · Hasnat Md. Abdullah[1] · Md. Golam Rabiul Alam[1] · Muhammad Mehedi Hassan[2] · Md. Zia Uddin[3]**

Md. Adyelullahil Mamun
md.adyelullahil.mamun@g.bracu.ac.bd

Hasnat Md. Abdullah
hasnat.md.abdullah@g.bracu.ac.bd

Md. Golam Rabiul Alam
rabiul.alam@bracu.ac.bd

Muhammad Mehedi Hassan
mmhassan@ksu.edu.sa

[1] Department of Computer Science and Engineering, BRAC University, Dhaka, Bangladesh

[2] Department of Information Systems, College of Computer and Information Sciences, King Saud University, Riyadh, Saudi Arabia

[3] Department of Software and Service Innovation, SINTEF Digital, Oslo, Norway

# Terms and Conditions

Springer Nature journal content, brought to you courtesy of Springer Nature Customer Service Center GmbH ("Springer Nature").

Springer Nature supports a reasonable amount of sharing of research papers by authors, subscribers and authorised users ("Users"), for small-scale personal, non-commercial use provided that all copyright, trade and service marks and other proprietary notices are maintained. By accessing, sharing, receiving or otherwise using the Springer Nature journal content you agree to these terms of use ("Terms"). For these purposes, Springer Nature considers academic use (by researchers and students) to be non-commercial.

These Terms are supplementary and will apply in addition to any applicable website terms and conditions, a relevant site licence or a personal subscription. These Terms will prevail over any conflict or ambiguity with regards to the relevant terms, a site licence or a personal subscription (to the extent of the conflict or ambiguity only). For Creative Commons-licensed articles, the terms of the Creative Commons license used will apply.

We collect and use personal data to provide access to the Springer Nature journal content. We may also use these personal data internally within ResearchGate and Springer Nature and as agreed share it, in an anonymised way, for purposes of tracking, analysis and reporting. We will not otherwise disclose your personal data outside the ResearchGate or the Springer Nature group of companies unless we have your permission as detailed in the Privacy Policy.

While Users may use the Springer Nature journal content for small scale, personal non-commercial use, it is important to note that Users may not:

1. use such content for the purpose of providing other users with access on a regular or large scale basis or as a means to circumvent access control;

2. use such content where to do so would be considered a criminal or statutory offence in any jurisdiction, or gives rise to civil liability, or is otherwise unlawful;

3. falsely or misleadingly imply or suggest endorsement, approval , sponsorship, or association unless explicitly agreed to by Springer Nature in writing;

4. use bots or other automated methods to access the content or redirect messages

5. override any security feature or exclusionary protocol; or

6. share the content in order to create substitute for Springer Nature products or services or a systematic database of Springer Nature journal content.

In line with the restriction against commercial use, Springer Nature does not permit the creation of a product or service that creates revenue, royalties, rent or income from our content or its inclusion as part of a paid for service or for other commercial gain. Springer Nature journal content cannot be used for inter-library loans and librarians may not upload Springer Nature journal content on a large scale into their, or any other, institutional repository.

These terms of use are reviewed regularly and may be amended at any time. Springer Nature is not obligated to publish any information or content on this website and may remove it or features or functionality at our sole discretion, at any time with or without notice. Springer Nature may revoke this licence to you at any time and remove access to any copies of the Springer Nature journal content which have been saved.

To the fullest extent permitted by law, Springer Nature makes no warranties, representations or guarantees to Users, either express or implied with respect to the Springer nature journal content and all parties disclaim and waive any implied warranties or warranties imposed by law, including merchantability or fitness for any particular purpose.

Please note that these rights do not automatically extend to content, data or other material published by Springer Nature that may be licensed from third parties.

If you would like to use or distribute our Springer Nature journal content to a wider audience or on a regular basis or in any other manner not expressly permitted by these Terms, please contact Springer Nature at

onlineservice@springernature.com